\documentclass[aps,prb,twocolumn,superscriptaddress,showpacs]{revtex4}
\usepackage{graphicx}
\usepackage[dvipdfm]{hyperref}
\begin{document}
\title{Spin-valve effect of the spin accumulation resistance in a double ferromagnet - superconductor  junction}
\author{P. S. Luo}
\affiliation{Institut N\'eel, C.N.R.S. and Universit$\acute{e}$ Joseph Fourier, 25 Avenue des Martyrs,
38042 Grenoble, France.}
\author{T. Crozes}
\affiliation{Institut N\'eel, C.N.R.S. and Universit$\acute{e}$ Joseph Fourier, 25 Avenue des Martyrs,
38042 Grenoble, France.}
\author{B. Gilles}
\affiliation{SIMAP, C.N.R.S., Universit$\acute{e}$ Joseph Fourier and Grenoble I.N.P., Domaine Universitaire, 1130 rue de la Piscine, 38402 Saint Martin d'H\`eres, France}
\author{S. Rajauria}
\affiliation{Institut N\'eel, C.N.R.S. and Universit$\acute{e}$ Joseph Fourier, 25 Avenue des Martyrs,
38042 Grenoble, France.}
\author{B. Pannetier}
\affiliation{Institut N\'eel, C.N.R.S. and Universit$\acute{e}$ Joseph Fourier, 25 Avenue des Martyrs,
38042 Grenoble, France.}
\author{H. Courtois}
\altaffiliation{Also at Institut Universitaire de France}
\affiliation{Institut N\'eel, C.N.R.S. and Universit$\acute{e}$ Joseph Fourier, 25 Avenue des Martyrs,
38042 Grenoble, France.}
\email[]{herve.courtois@grenoble.cnrs.fr}
\date{\today}

\begin{abstract}
We have measured the transport properties of Ferromagnet - Superconductor nanostructures, where two superconducting aluminum (Al) electrodes are connected through two ferromagnetic iron (Fe) ellipsoids in parallel. We find that, below the superconducting critical temperature of Al, the resistance depends on the relative alignment of the ferromagnets' magnetization. This spin-valve effect is analyzed in terms of spin accumulation in the superconducting electrode submitted to inverse proximity effect.
\end{abstract}

\pacs{74.45.+c}
\maketitle

At a Normal metal-Superconductor (N-S) junction biased at a voltage below the superconducting gap $\Delta/e$, Andreev reflection is the dominant contribution to transport. Here, one spin-up electron penetrates the superconductor with a spin-down electron so that a Cooper pair is formed. This can also be viewed as the reflection of an electron into a hole.\cite{Andreev} Since the two electron spin populations are involved, Andreev reflection is reduced in a F-S junction based on a Ferromagnetic (F) metal, and suppressed in the case of a full spin-polarization.\cite{PRL-Beenakker}

Electron transport in F-S hybrid structures exhibits several remarkable phenomena. Crossed Andreev Reflection (CAR) is predicted to occur when two normal metal leads contacting a superconductor are separated by at most the superconducting coherence length $\xi_s$:\cite{PRL-Byers} one electron from one lead is reflected as a hole in the other lead. Similarly, an electron can travel from one lead to the other by elastic co-tunneling (EC).\cite{EPL-Falci} Experiments on multi-terminal F-S structures showed a non-local signal sensitive to the relative magnetization alignment.\cite{PRL-Beckmann} This was attributed to CAR, which is enhanced in a anti-parallel state (AP) and inhibited in a parallel (P) state. A similar bias-dependent non-local signal was observed in N-I-S-I-N planar junctions (I stands for Insulator)\cite{PRL-Russo} and N-S multerminal devices.\cite{PRL-Chandra} Spin switches made of one superconductor sandwiched by two ferromagnetic layers have been studied close to the superconductivity critical temperature $T_c$, which was found to be larger in the AP state than in the P state\cite{PRL-Gu} in accordance to the expected proximity effect.\cite{PL-deGennes} Opposite behaviors were also observed\cite{PRB-Rusanov} and explained in terms of stray field effects.\cite{PRB-Steiner} At a F-S junction, a spin-polarized current is converted into a spinless current.\cite{PRB-vanWees} This occurs in the ferromagnetic metal on a characteristic length scale given by the spin relaxation length $\lambda_{sf}$. Electrons with the minority spin then accumulate close to the interface, which induces an extra resistance of amplitude determined by a length $\lambda_{sf}$. In a F-I-S tunnel junction, quasi-particules can be injected only at an energy above the gap, generating a large spin-accumulation resistance.\cite{APL-Johnson,PRL-Haviland}

Although non-local mechanisms, spin switch and spin accumulation effects can coexist in hybrid nanostructures, their relative contribution to electron transport has been little studied. In this paper, we address the spin-dependent transport at the junction between two ferromagnetic leads and a superconductor, in the regime of a metallic contact. We observe a bias-dependent spin-valve effect, which we analyze in terms of spin accumulation in the superconducting electrode submitted to inverse proximity effect.

\begin{figure}[t]
\includegraphics[width = 8 cm]{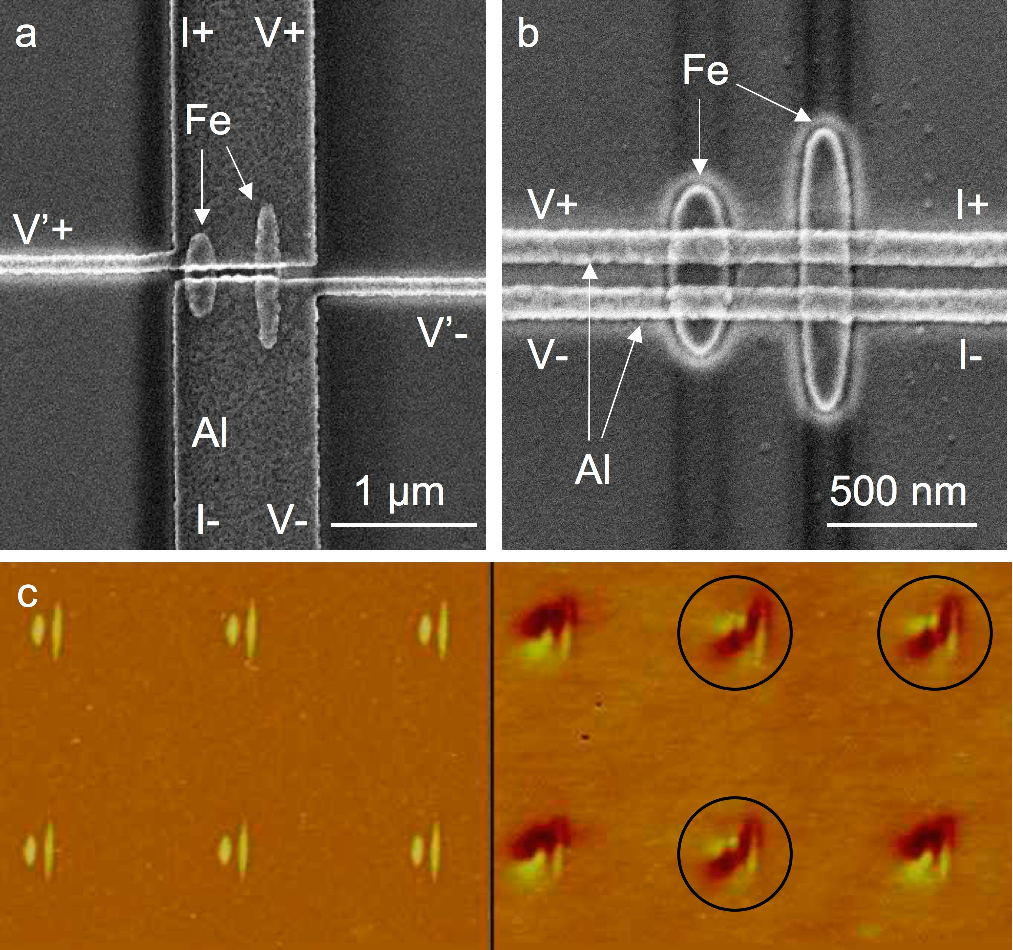}
\caption{(Color online) Top: Micrographs of the two sample geometries based on a Fe ellipsoids pair together with the measurement connections. (a) The two wide Al pads have voltage probes close to the interface. (b) The two Al wires do not overlap the ends of the ellipsoids, where the largest stray field is induced. Bottom: (c) Topographical (left) and magnetic (right) images of the same area of a test sample made of a large number of Fe ellipsoids pairs. The magnetic image was taken at a magnetic field of 30 mT after having polarized the sample in the opposite direction at  - 200 mT. The pairs indicated by a circle show an anti-parallel (AP) magnetization state.}
\label{sample}
\label{MFM}
\end{figure}

Fig. \ref{sample} a, b show the two sample geometries that we have investigated. In every case, two superconducting Al reservoirs or wires are connected through two Fe ellipsoids in parallel. We have chosen an ellipsoidal shape in order to ensure a single magnetic domain regime within one ellipsoid. The spacing between the Fe ellipsoids was varied between 100 and 500 nm. The separation between the two Al reservoirs is 100 nm, which is much larger than the proximity effect decay length in a ferromagnetic metal. This means that the two superconducting interfaces of the same ellipsoid are decoupled. Geometry a) is designed to have bulk Al contacts with voltage probes close to the interface, while geometry b) reduces significantly the influence on the Al electrodes of the stray field induced at the Fe ellipsoids ends. The fabrication procedure starts from epitaxial Fe films that were grown on a MgO substrate at room temperature under a residual pressure below 10$^{-9}$ mbar and annealed at 600 $^{\circ}$C for 3 h. The films are 40 nm thick and protected by a 3 nm layer of Pt or Au. First, the Fe ellipsoids are patterned by e-beam lithography and Ar ion-etching. After a second e-beam lithography, a 70 nm Al film is deposited on a resist mask and lifted-off. Prior to the deposition of Al, the protection layer is removed by a soft ion-milling.

In a given sample, the two ellipsoids have been made with different dimensions (900 $\times$100 nm$^2$ and 500 $\times$150 nm$^2$) in order to obtain different coercive fields. Fig. \ref{MFM}c displays the topographical and the magnetic images of the same area of a test sample featuring a large number of Fe ellipsoids pairs. The magnetic images were acquired with a Magnetic Force Microscope (MFM) and give access to the perpendicular to the surface component of the magnetic field gradient. Fig. \ref{MFM}c data was acquired at a moderate in-plane magnetic field of 30 mT after full polarization of the sample in the opposite direction at - 200 mT. For every ellipsoid, the magnetic image is compatible with a single magnetic domain configuration. Although all ellipsoids pairs were made identically, part of them show an AP magnetization configuration, meaning that the short ellipsoid has switched, while the long one remains pinned. This scattering is presumably due to slight changes in the precise ellipsoids geometry. For instance, it is expected that the ellipsoid edge roughness plays a significant role in the exact value of the coercive field. Based on the full series of measurements, we find that the switching fields of the ellipsoids are about 30 mT for the short one and 50 mT for the long one, with significant variations from one sample to the other.

We have measured the electron transport properties of a series of samples at very low temperature down to 260 mK. We used a lock-in technique with an \mbox{a.c.} current of amplitude 100 to 200 nA superposed to a \mbox{d.c.} bias current. Here we present experimental data from one sample (Sample 5) out of 8 samples showing a similar behavior. It is of geometry a) and was measured in a 2-wire configuration, unless otherwise specified. The data displayed here was acquired at zero applied magnetic field.

Let us first discuss the spin-valve effects in our samples. First we polarized the two ellipsoids magnetizations with a magnetic field of absolute value 300 mT. Afterwards and for every data point, we applied for about 1 s a probe magnetic field of a varying value. The field is then ramped back to zero and the resistance is measured. This procedure is repeated at a series of values for the probe magnetic field, starting from the polarization field and until a field opposite in sign is reached. In this way, we systematically measure the resistance of the device in different magnetization configurations, without the parasitic effects of a non-zero magnetic field.

\begin{figure}[t]
\includegraphics[width = 8 cm]{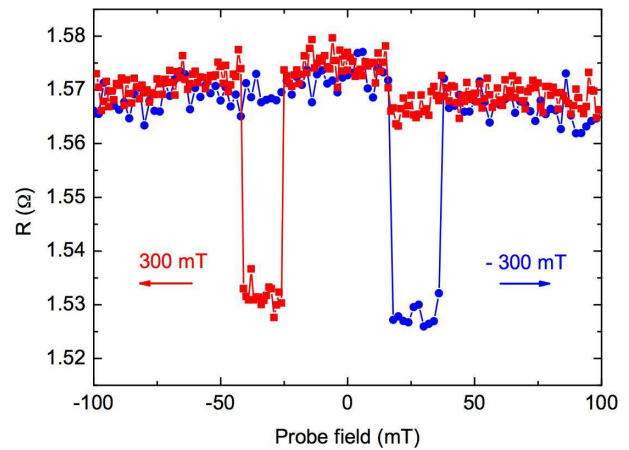}
\caption{(Color online) Probe magnetic field dependence of Sample 5 zero-bias differential resistance at 310 mK. Red square (blue circular) dots: the Fe ellipsoids are first polarized with + (-) 300 mT magnetic field and the differential resistance is measured at zero external magnetic field with decreasing (increasing) probe fields. \label{RPH}}
\end{figure}

Fig. \ref{RPH} shows an example of such a measurement, where the horizontal axis indicates the probe magnetic field that was applied just before the measurement. We observe sharp stepwise changes of the resistance, with two symmetric domains featuring a lower value. The values of the probe magnetic field at the resistance changes are compatible with the switching fields of the two different ellipsoids. In this respect, we ascribe the low-resistance domains to the regime of AP magnetizations ($\uparrow\downarrow$ or $\downarrow\uparrow$). Both at smaller field and at higher field, the resistance is higher and constant within the measurement accuracy. We ascribe these states to a P configuration ($\downarrow\downarrow$ or $\uparrow\uparrow$). The resistance difference between the AP and the P states is about 40 m$\Omega$ or 3 $\%$. This spin-valve effect is the central result of this paper. Let us now describe our further experimental study and data analysis aimed at identifying the involved physical effect.

\begin{figure}[t]
\includegraphics[width = 8 cm]{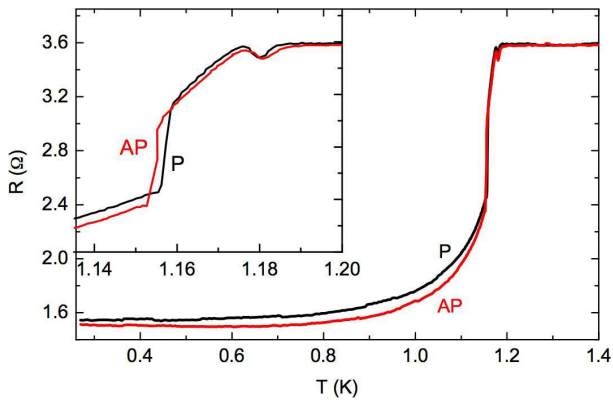}
\caption{(Color online) Temperature dependence of Sample 5 zero-bias resistance in parallel (P) and anti-parallel (AP) magnetization states. The inset shows a zoom close to the Al superconducting critical temperature.\label{RT}}
\end{figure}

Fig. \ref{RT} inset shows Sample 5 resistance temperature dependence, when prepared in a P or AP state. The magnetic state has a small effect (about 3 mK) on the critical temperature of the Al electrodes. The sign of the shift (at the sharpest resistance drop) suggests an effect of the dipolar magnetic field arising from the Fe ellipsoids.\cite{PRB-Steiner} Some other samples showed an opposite effect but with a similar amplitude, compatible with a dominating proximity effect.\cite{PL-deGennes} At low temperature, this spin switch effect may modify the superconducting gap and hence influence the transport properties. Nevertheless, all samples showed a similar spin-valve behavior although they exhibit a different spin switch effect. Moreover, the voltage across the device is well below the gap in the discussed data. Thus the spin switch effect does not explain the spin-valve behavior observed at very low temperature. 

\begin{figure}[t]
\includegraphics[width = 6 cm]{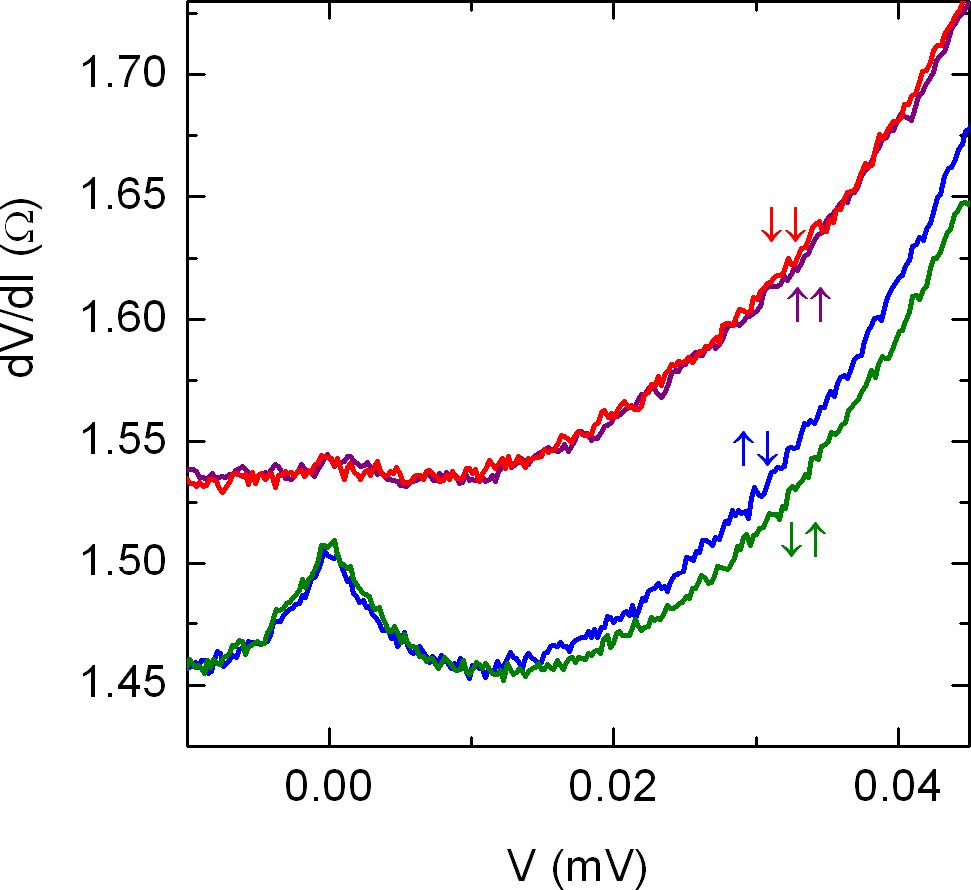}
\includegraphics[width = 2.2 cm]{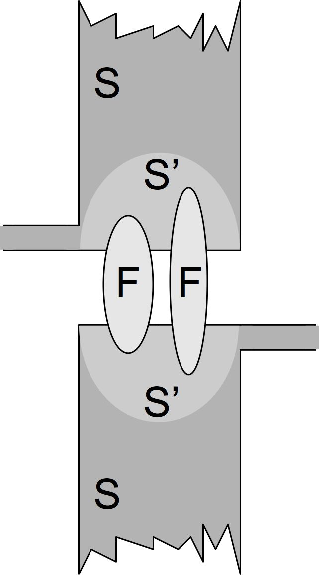}
\caption{(Color online) Left: Voltage dependence of Sample 5 differential resistance in the two P (top curves, $\downarrow\downarrow$ and $\uparrow\uparrow$) and the two AP states (bottom, $\uparrow\downarrow$ and $\downarrow\uparrow$). Right: Schematics of the sample geometry a) outlining the region S' of the superconducting electrodes S submitted to inverse proximity effect and where spin accumulation is expected to occur.\label{dIdV}}
\end{figure}

Fig. \ref{RT} main panel displays the Sample 5 resistance temperature dependence in both P and AP states. Below $T_c$, the resistances in the two states show a non-monotonous behaviour. The spin-valve effect amplitude has also a non-monotonous behavior, with a maximum at about 0.9 K. At lower temperature, it decreases steadily towards zero. The effect is absent within a 1 m$\Omega$ resolution above the critical temperature $T_c$ = 1.18 K of Al. This confirms that the observed spin-valve effect is related to superconductivity. Fig. \ref{dIdV} left shows Sample 5 differential resistance in different magnetic states. In the AP states, we find a zero-bias resistance peak, which is suppressed in the P states. The spin valve effect is larger for finite voltage bias, which is consistent with the above observation that it is larger at finite temperature. The two P states on one side and the two AP states on the other side behave very similarly. This confirms that electron transport depends on the relative magnetization alignment of the two ellipsoids, not on the direction of a given magnetization. 

Table 1 lists the main properties of the 8 investigated samples, showing a spin-valve effect as discussed above, at 275 mK. The spin-valve effect amplitude $\Delta R$ varies quite little, between 18 to 43 m$\Omega$. The sample resistance $R$ is small and quite constant for Samples 2-5, where all F-S interfaces are very transparent. Samples 1 and 6-8 with a larger resistance feature presumably less transparent F-S interfaces at one of the superconducting electrodes, bringing to almost zero the spin-valve effect in that electrode. In the geometry a), we have probed the resistance either in a 2 wires geometry by using the wide Al electrodes for both current bias and voltage measurement, or in a 4 wires geometry by using the voltage probes for the measurement. We observed an identical low temperature behavior in both cases. In a 4-wire configuration, a resistance peak appears close to the critical temperature, due to charge-imbalance in the Al pads.\cite{PRL-Clarke}

Let us now turn to the interpretation of the observed spin-valve effect. As for non-local effects, EC would have no contribution here, since we current-bias the two ellipsoids in parallel. The sign and amplitude of the measured effect are compatible with CAR. Nevertheless, the absence of a significant influence of the ellipsoids separation in the 100-500 nm range investigated here, whereas the coherence length $\xi_s$ is estimated to be about 100 nm, discards an interpretation in terms of CAR. 

A significant inverse proximity effect is expected in a superconductor in metallic contact with a ferromagnetic metal. The sub-gap electronic density of states is non-zero in a region S' extending in the superconductor over a few times the coherence length $\xi_s$,\cite{EPL-Sillanpaa} see Fig. \ref{dIdV} right part. This allows for the injection, even at a sub-gap bias, of spin-polarized quasi-particules from the two Fe ellipsoids into every Al electrode. In a P state, the current through both ellipsoids injects the same majority of spins and a significant spin accumulation builds up in S'. In a AP state, one ellipsoid injects spin-up quasi-particles and the other one injects spin-down quasi-particles. The two spin populations are then balanced and little spin accumulation is expected in S'. Thus a AP state is expected to have a lower resistance than a P state, as observed in the experiment.

\begin{table}
\begin{ruledtabular}
\begin{tabular}{ccccccccc}
Sample & Geometry & Separation (nm) & $R (\Omega)$ & $\Delta R (\Omega)$\\
\hline
1 & a & 150 & 9.21 & 0.022\\  
2 & a & 150 & 1.94 & 0.041\\ 
3 & a & 150 & 2.16 & 0.023\\ 
4 & b & 100 & 2.76 & 0.018\\ 
5 & a & 150 & 1.54 & 0.043\\ 
6 & b & 150 & 8.37 & 0.024\\ 
7 & b & 150 & 18.45 & 0.018\\ 
8 & b & 500 & 5.62 & 0.023\\ 
\end{tabular}
\end{ruledtabular}
\caption{Sample parameters including the geometry type, the ellipsoids separation, the P state resistance $R$ and the spin-valve effect amplitude $\Delta R$ both at 275 mK. \label{Table}}
\end{table}

The considered spin-accumulation builds up within the Al electrode in the region S', while the Andreev reflection occurs at the S'-S interface. No spin-valve effect is expected in the absence of inverse proximity effect, in which case spin accumulation would occur separately in the two ellipsoids. The observation of a spin-valve effect in Sample 8 with a ellipsoids separation of 500 nm indicates that the region S' extends over about 250 nm from a F-S interface, which is less than 3 times the coherence length $\xi_s$. This is in agreement with Ref. \onlinecite{EPL-Sillanpaa}, which shows that a significant sub-gap density of states level remains at such a distance. The inverse proximity effect and the related spin accumulation thus decay slower than the CAR amplitude\cite{PRL-Beckmann,PRL-Russo} when the separation between the two ellipsoids is increased in the range of a few times the superconducting coherence length.
  
At an F-S interface, the magnitude of the spin accumulation induced resistance is $\Delta R = R_{sq}.(\lambda_{sf}/w).(\alpha^2/(1-\alpha^2))$, where $R_{sq}$ is the square resistance, $w$ the wire width and $\alpha$ is the spin polarization.\cite{PRB-vanWees} Applying this analysis to the S'-S interface, with a square resistance $R_{sq}$ of 1 $\Omega$, a wire width $w$ of 400 nm, a spin relaxation length $\lambda_{sf}$ of 400 nm in Al \cite{PRB-vanWees-2} and a polarization $\alpha$ of 40 $\%$ close to the one of bulk Fe,\cite{PRL-Mazin} one obtains $\Delta R$ = 0.02 $\Omega$ per interface, in fair agreement with the resistance change amplitude at zero bias. The spin polarization in the Al electrode S' region is presumably smaller than in bulk Fe, which would decrease the amplitude of the effect. Eventually, the dependence on bias and temperature of the spin-valve effect can be related to changes of the size of the region S'. As the current bias or the temperature increases, the inverse proximity effect extends over a larger distance and the spin accumulation effects increase.

In conclusion, we have investigated the sub-gap transport properties in double F-S hybrid structures with two F elements. Below the critical temperature of the superconductor, the resistance depends on the relative magnetization alignment. This spin-valve behavior is related to the spin accumulation in the superconducting electrode submitted to inverse proximity effect. This approach is similar to considering a out-of-equilibrium region in the vicinity of the interface \cite{PRB-Melin} and may hold for previous experiments in similar hybrid structures.\cite{EPJB-Giroud}

The samples have been fabricated at Nanofab-C.N.R.S. Grenoble plat-form. We thank M. Giroud for contributing at the early stage of this project, I. L. Prejbeanu for the MFM measurements, D. Beckmann and R. M\'elin for discussions. We acknowledge support from "Elec-EPR" ANR contract and STREP "SFINx" EU project.

\end{document}